# Dedicated SiPM array for GRD of GECAM

D.L. Zhang[1], X.L. Sun[*1,2], Z.H. An[*1], X.Q. Li[1], X.Y. Wen[1], K. Gong[1], C. Cai[1], Z. Chang[1], G. Chen[1], C. Chen[1], Y.Y. Du[1], M. Gao[1], R. Gao[1], D.Y. Guo[1], J.J. He[1], D.J. Hou[1], Y.G. Li[1], C.Y. Li[1], G. Li[1], L. Li[1], X.F. Li[1], M.S. Li[1], X.H. Liang[1], X.J. Liu[1], Y.Q. Liu[1], F.J. Lu[1], H. Lu[1], B. Meng[1], W.X. Peng[1], F. Shi[1], H. Wang[1], J.Z. Wang[1], Y.S. Wang[1], H.Z. Wang[1], X. Wen[1], S. Xiao[1], S.L. Xiong[1], Y.B. Xu[1], Y.P. Xu[1], S. Yang[1], J.W. Yang[1], Fan Zhang[1], S.N. Zhang[1], C.Y. Zhang[1], C.M. Zhang[1], Fei Zhang[1], X.Y. Zhao[1], X. Zhou[1]

[1] *Institute of High Energy Physics, CAS, Beijing 100049, China;*
[2] *State Key Laboratory of Particle Detection and Electronics. Beijing 100049, China;*
\* E-mail: sunxl@ihep.ac.cn, anzh@ihep.ac.cn

**Abstract**

The discovery of gravitational waves and gamma-ray bursts heralds the era of multi-messenger astronomy. With the adoption of two small satellites to achieve the all-sky monitoring of gamma-ray bursts, the gravitational wave high-energy electromagnetic counterpart all-sky monitor (GECAM) possesses a quasi-real-time early warning ability and plays an important role in positioning the sources of gravitational waves and in subsequent observations. Each satellite of GECAM was fitted with 25 3-inch-diameter gamma-ray detectors (GRD), covering an energy range of 8 keV–2 MeV. GRDs have adopted silicon photomultiplier tubes (SiPM) in lieu of photomultiplier tubes (PMT) to adapt to the dimensional limitations of micro-satellites and used lanthanum bromide crystals with a high light yield in order to lower the threshold to 8 keV. In this study, a unique 3-inch circular SiPM array was designed. In this design, 64 6x6 mm chips were arranged evenly in a circular manner with the seams filled with reflecting films, thus achieving satisfactory uniformity of light collection. The integrated pre-amplifier circuit on the back of the SiPM array adopted two-level grouping and summing; further, it achieved a satisfactory signal-to-noise ratio. Two high-gain and low-gain channels were adopted to achieve a large dynamic range, and two independent power supply units were used, where each unit can be closed separately, thus improving reliability. In this article, the structural design, performance, and comparison with PMTs are introduced.

## 1. Introduction

The discovery of gravitational waves and gamma-ray bursts heralds the era of multi-messenger astronomy [1][2]. However, there is a lack of specialized scientific satellites that can detect the electromagnetic counterparts of gravitational waves, meaning all existing space-based detectors have drawbacks such as a small field of view and poor positioning accuracy. This results in difficulties when coordinating the search for rare gravitational events with detections from LIGO/VIGO. To address these limitations, the gravitational wave high-energy electromagnetic counterpart all-sky monitor (GECAM) utilizes two small satellites positioned on opposite sides of the Earth for an all-sky search of gamma-ray bursts and quasi-real-time download of trigger signals, thus becoming an important space-based detector in the era of multi-messenger astronomy [3]. Small satellites have the benefits of a short time cycle and low cost. However, this is a challenge for the design of detectors, as there are restrictions on size, weight, and power consumption.

Therefore, for GECAM, detectors with a compact structure and satisfactory performance were developed by substituting traditional photomultiplier tubes (PMT) with silicon photomultiplier tubes (SiPM), together with lanthanum bromide crystals with a high light yield [4][5][6]. As a result, each 46.5-cm-diameter small satellite was arranged with over 25 3-inch gamma-ray detectors (GRDs) and eight 2-inch charge particle detectors. The large array of detectors ensures the sensitivity and positioning accuracy of the GECAM. As semiconductor photoelectronic devices, SiPMs have undergone rapid development in recent years with significant improvements in performance. In particular, their photon detection efficiency (PDE) reached 50% [7], which is 40% higher than that of traditional high-quantum-efficiency PMTs. Therefore, scintillation detectors have been increasingly adopted in high-energy physical experiments [8][9][10] and nuclear medicine [11]. Furthermore, several elements of space detectors have also adopted SiPMs [12][13]. In this study, a 3-inch SiPM array with a unique structure was designed for the GRDs of GECAM, and the main design considerations, performance, and feasibility of popularization and application are discussed.

**2. Design of SiPM array**

The design of the 3-inch SiPM array predominantly comprises the selection of SiPMs, the quantity and arrangement of chips, and the pre-amplifier electronics. The main considerations for the selection of SiPMs are spectral fitness with lanthanum bromide crystals, PDE, dark count rate (DCR), crosstalk, pixel and chip size, sealing, consistency, operating bias voltage, irradiation resistance, etc. The main considerations for the quantity and arrangement of chips in an array are the signal-to-noise ratio and the evenness of the detected photons and DCR. The main considerations for the design of the pre-amplifier circuit are the signal-to-noise ratio, reliability, dynamic range, compensation of temperature effect, and suitability with a data acquisition system.

The SiPM chips chosen for this study were SensL's J-series SiPMs, with a 5 V bias voltage PDE decreases from 40 % to 50 % given that the luminous peak wavelength of lanthanum bromide is 380 nm and the DCR is less than 70 kHz/mm$^2$ at 21 °C. To reduce the number of routing channels, the size of the chips was selected to be 6x6 mm, which is the largest among mass-produced products. The pixel size is 35 μm, which has a better linear dynamic range than 50 μm or larger pixels, and the TSV package has a high fill factor. In particular, the SensL SiPMs have a better bias voltage consistency than that of similar products from Hamamatsu and FBK, which is important for building large-area arrays. Therefore, we can use the same supply voltage as we would for one SiPM array, which simplifies the circuit design. Furthermore, the bias voltage is less than 30 V, which matches the power supply of the small satellite platform.

Within the 3-inch diameter area, a ring structure was adopted for the arrangement of 64 SiPM chips with a filling rate of 50 %. The light yield of the lanthanum bromide was 63 photons/keV γ, and the decay time was 16 ns. At 5 keV, the light yield was 47 photons/keV, which was relatively low because of the nonlinearity of the crystals. However, 5 keV X-rays can still excite up to 236 photons within 100 ns. Considering the fact that the photon collection efficiency is 50 % and the PDE of the SiPMs is 40 % with a bias voltage of 5 V, the resulting photoelectron number is 47 p.e. The corresponding thermal noise of the SiPMs was approximately 16 p.e. within 100 ns, so the ratio of the number of photoelectrons to the signal to noise (signal-to-noise ratio) is 2.9. To improve the photon collection efficiency, the seams of the SiPM chips were filled with Tyvek reflecting films with a reflection rate of >95 % at 380 nm. Given the 50 % coverage of the Tyvek reflecting films, and without considering the self-absorption of crystals, it can be calculated that the photon collection efficiency was improved to 95 %, and the signal-to-noise ratio increased to 5.6. To further improve the signal-to-noise ratio, the temperature of the detector on the satellite is set to -20 °C using a thermal control system, and the resulting thermal noise can be reduced by nearly an order of magnitude. Therefore, the design meets the requirement of GECAM to detect 8 keV low-energy X-rays. Generally, there are two different schemes for SiPM front-end electronics; one is in the form of discrete devices, and the other is in the form of ASIC chips. From the perspective of space application reliability, we chose a discrete device solution. Because SiPM is a capacitive device, it is necessary to aggregate 64 pieces into one channel by grouping and summing, instead of directly connecting 64 pieces into one channel in parallel. We therefore divide the 64 SiPMs into four groups where each group of 16 pieces is connected in parallel and the first level summation is performed, and the four groups are then connected in parallel for the second level summation. To meet the requirements of the 8 keV - 2 MeV dynamic range, we divide the output of the summing circuit into high-gain and low-gain channels. The high-gain channel is used for low-energy X-ray detection and its output is connected to the data acquisition system through a differential circuit, while the low-gain channel is used for high-energy gamma-ray detection, where the signal is relatively large and does not use a differential circuit in the data acquisition system.

We used four pieces of dual high-speed voltage feedback amplifier chips, a total of eight channels of operational amplifiers, six channels for high- and low-gain summing circuits, and two channels were used for the differential output. The conceptual design of the summing circuit is shown in Figure 1. In particular, the arrays were provided with two independent power supply units due to the possibility of power supply failure due to the short circuit of individual SiPMs. Therefore, if one unit is closed off, the other unit will still provide a power supply, and only half of the signal amplitude will be reduced.

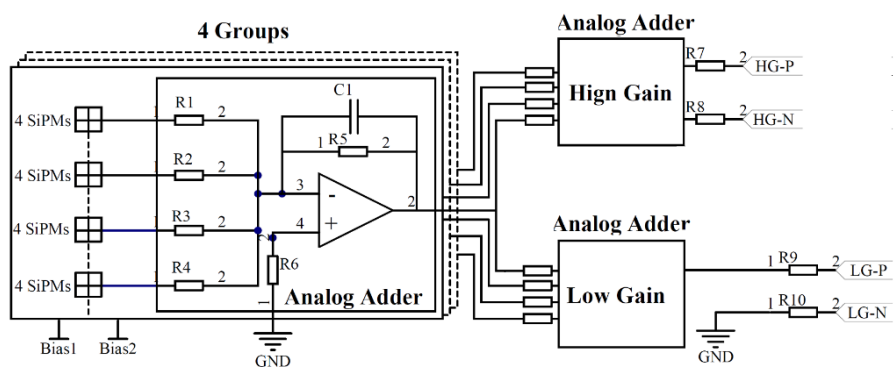

Figure 1. The conceptual design of the summing circuit.

SiPMs are temperature-sensitive devices, and therefore the dark count rate and the breakdown voltage are regulated by temperature. One temperature measurement chip was consequently arranged on the back side of the SiPM array circuit board to monitor the ambient temperature of each array. The temperature data were transmitted to a digital management system which would then adjust the bias voltage of the SiPM power supply, based on the breakdown voltage change rate of approximately 21 mV/°C, thus ensuring predominantly consistent gains of the SiPMs at different temperatures. Ultimately, temperature feedback is only one of the adjustment methods for detector consistency. The fundamental consistency calibration method is to perform on-orbit calibration through the two characteristic peaks of lanthanum bromide, which are 37.4 keV and 1.4 MeV.

Based on the above design considerations, a 3-inch diameter SiPM array for GECAM GRD was designed, as shown in Figure 2. Subsequently, a performance test was conducted.

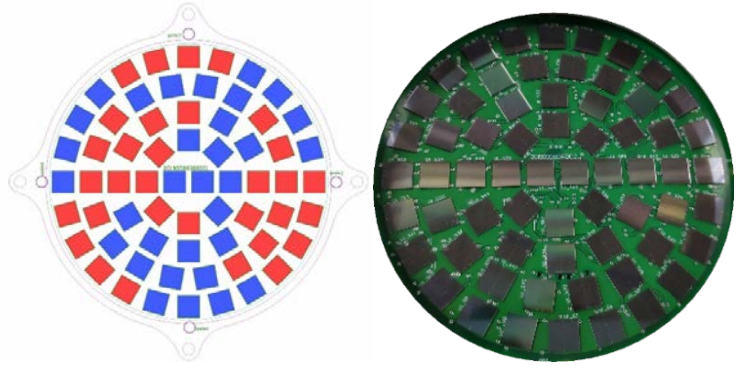

Figure 2. The layout of SiPM array and power supply. The red and yellow indicate two groups of power supply respectively (left). The photo of SiPM array (right).

### 3. Test setup for the SiPM array

After the SiPM array was coupled with lanthanum bromide crystals using silicone pads, a complete GRD was assembled, as shown in Figure 3. The lanthanum bromide crystals were domestic products, the silicone pads were from Saint-Gobain with thickness of 1 mm, and the Tyvek reflecting films were from DuPont. The high-gain and low-gain signals are transmitted through a multi-channel MCA8000A for data acquisition after formation by the main amplifier ORTEC671. The radioactive sources of $^{55}$Fe, $^{241}$Am, and $^{137}$Cs were adopted to measure the energy spectrum and energy resolution, and a collimated X-ray source was used for the measurement of uniformity. A low-temperature performance test was carried out by placing the entire GRD into a low-temperature experimental box with a temperature control accuracy of 0.1 °C. In contrast to PMTs, Hamamatsu R6233-100 with a high quantum efficiency was adopted. The test of temperature feedback regulation will be discussed in another article about the 100 m beamline calibration experiment, and calibrations of the energy-count relations will be detailed in the calibration article.

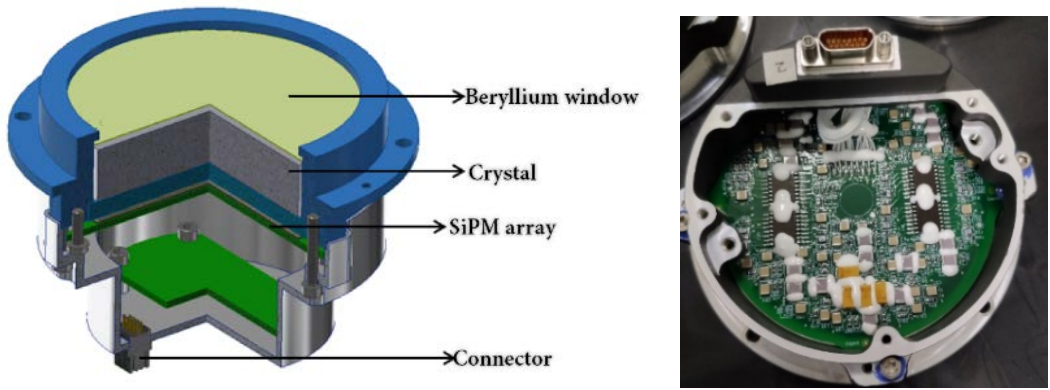

Figure 3. A schematic diagram of detector design (left) and a photo of the GRD back side (right).

### 4. Performance

At room temperature, the waveforms of typical high-gain and low-gain X-rays that are not formed through the main amplifier are shown in Figure 4. The amplitude of the high-gain signals was four times that of the low-gain signals, with a bottom width of approximately 2 μs and a front rise time of approximately 150 ns. High-gain signals were mainly used for subsequent tests and analyses. The energy spectrum collected through multiple channels after passing through the main amplifier is shown in Figure 5. The resolution of the 662 keV [137]Cs was 3.78 %, and that of the 59.5 keV [241]Am was 13.5 %. Comparatively, the resolutions of the PMTs were 3.85 % and 11.2 %, respectively. The resolution of the 662 keV PMT was relatively poor, and that of the 59.5 keV PMT was comparatively better.

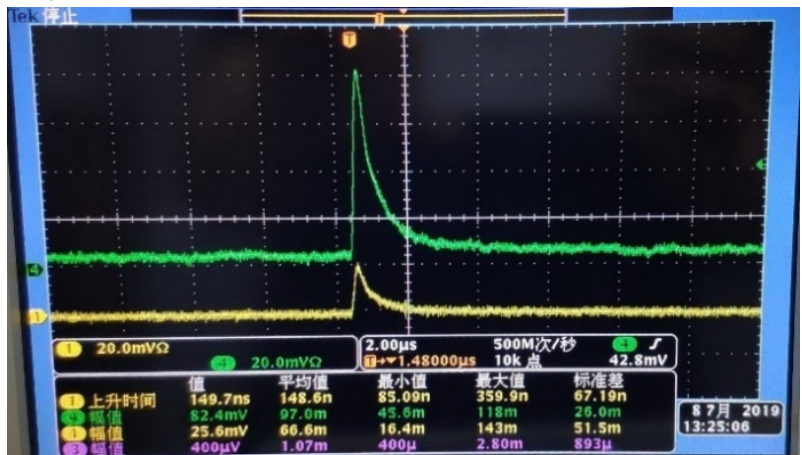

Figure 4. Typical waveforms of high- and low-gains.

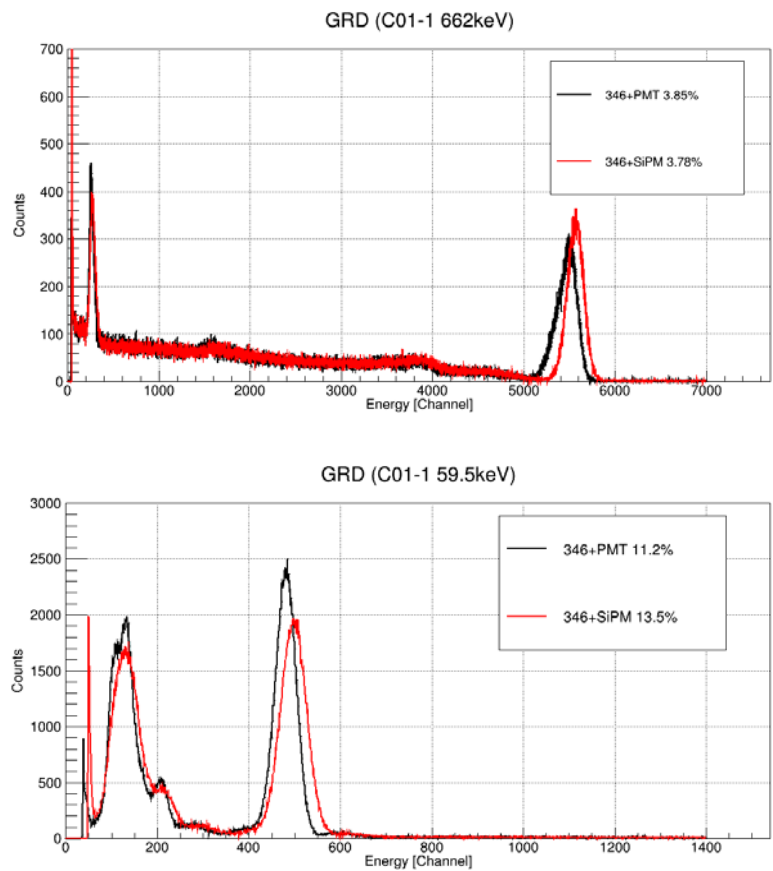

Figure 5. The measured energy spectrum of SiPMs and PMTs excited by [137]Cs source (top) and [241]Am source (bottom) at room temperature.

As the temperature decreases, the breakdown voltage decreases, but here we adopt the form of a fixed bias voltage and therefore the bias voltage does not decrease with temperature. The change in the 59.5 keV peak and energy resolution are shown in Figure 6. It is obvious that the peak rises and the energy resolution improves with a decrease in temperature, with a rapid change from 30 °C to 10 °C, and a slower change from 10 °C to -30 °C. The increase in peak position is due to the decrease in the breakdown voltage of SiPMs as the temperature decreases, and the fixed bias voltage is equivalent to the increase in the overvoltage (bias voltage–breakdown voltage) of SiPMs. The main contribution of energy resolution comes from the Poisson component of photons to photoelectrons, the fluctuation of the photon production of the crystals, electronic noise, and DCR of the SiPMs. When the DCR contribution is significantly less than the other contributions, the energy resolution is dominated by other factors that are weakly related to temperature. Thus, the platform of the energy resolution curve can appear when the temperature is sufficiently low. The 5.9 keV X-ray energy spectrum of $^{55}$Fe at a low temperature of -30 °C is shown in Figure 7. The energy resolution of the full-energy peak was 42.4 %.

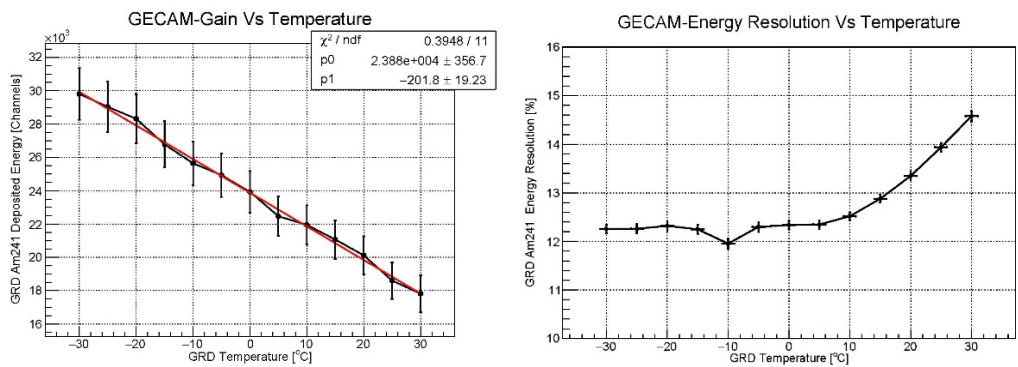

Figure 6. Change of 59.5 keV X-ray peak(left) and energy resolution(right) of $^{241}$Am source with change of temperature.

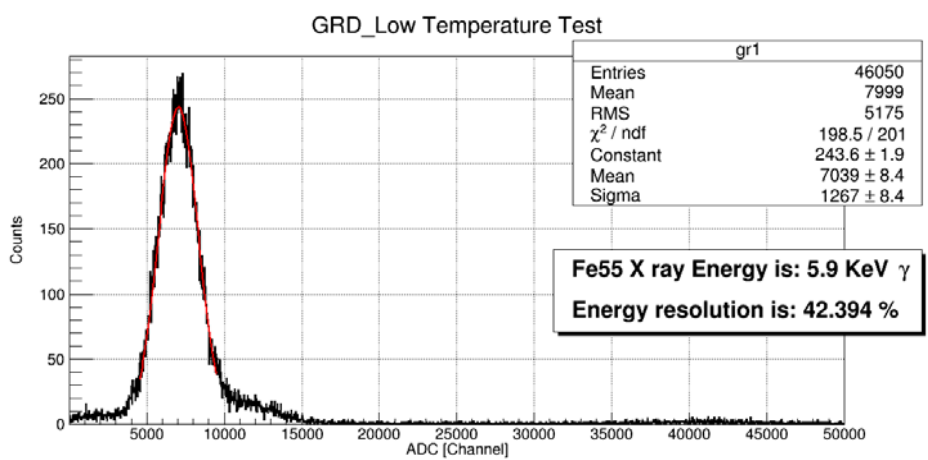

Figure 7. Energy spectrum of 5.9 keV $^{55}$Fe X-ray at low temperature (-30 °C).

The uniformity measurement points are shown in Figure 8. An incident point was selected at intervals of 11 mm across the diameter, and a total of seven test points were selected. The X-ray energy was 13.3 keV, 64.1 keV, 83.6 keV and 130.1 keV respectively. The results suggested that the position uniformity of all X-rays with different energies was better than 2 %, indicating satisfactory consistency of SiPMs and uniformity of light collection.

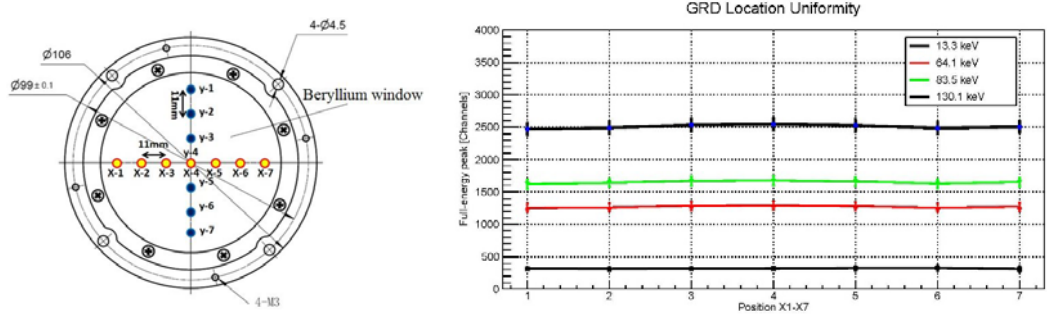

Figure 8. Uniformity test, position of vertical incidence of X-ray (left), change of peak value along the diameter position (right).

Three X-ray energy points of 25.9 keV, 64.1 keV and 130.1 keV were adopted to test the half-unit power supply, where the energy spectrum is shown in Figure 9, and the data is shown in Table 1. The difference in the peaks of the two half-units was smaller than 5 %. Because there was a reduced number of photons received with the half-unit power supply, the energy resolution decreased accordingly. Interestingly, the proportion of the peak values retrieved from the half-unit power supply compared to that from the whole-unit was not quite the same for different energy points, and the proportions were 61.7 %, 51.4 %, and 49.7% from low energy to high energy, respectively. It is clear that the lower the energy, the higher the proportion of the half-unit. The reason for this result is still unclear, and further research will be conducted in the future.

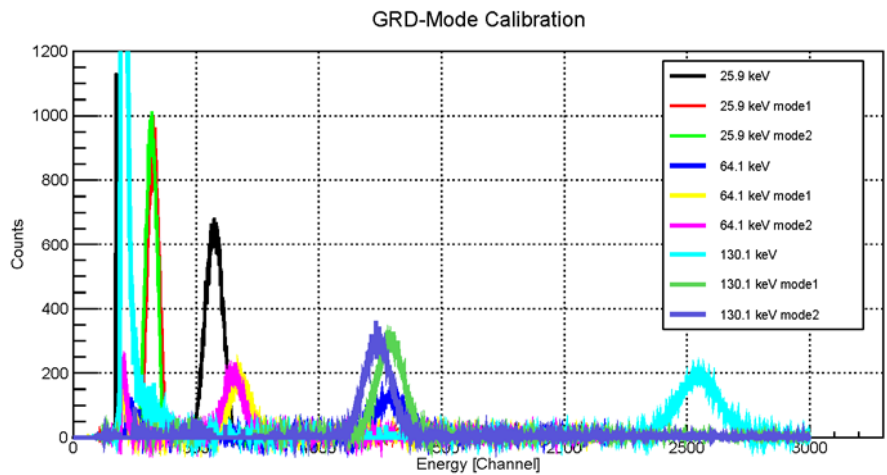

Figure 9. Energy spectrum of X-ray under whole-unit and half-unit power supply.

Table 1. Test data under whole-unit and half-unit mode.

| Energy keV | mode | mean | sigma | Resolution |
|---|---|---|---|---|
| **25.9** | Half 1 | 325.6 | 23.0 | 16.6 |
| | Half 2 | 317.9 | 22.3 | 16.5 |
| | Whole | 527.6 | 35.6 | 14.6 |
| **64.1** | Half 1 | 677.7 | 41.7 | 14.5 |
| | Half 2 | 651.6 | 39.9 | 14.4 |
| | Whole | 1293.0 | 63.4 | 11.5 |
| **130.1** | Half 1 | 1292.0 | 55.7 | 10.1 |
| | Half 2 | 1239.0 | 55.3 | 10.1 |
| | Whole | 2548.0 | 88.5 | 8.1 |

## 5. Conclusion and discussion

In response to the requirements that GECAM should have a compact structure, low threshold, and a large GRD dynamic range, we have developed a 3-inch SiPM array that uses grouped summation, reflective films, a circular arrangement, two groups of independent power supplies, high- and low-gain signals, differential signal output technologies, etc. This solution can be used not only for GECAM, but also as a general solution for SiPM-based scintillation detectors. The design herein realized the single-channel reading of the SiPM array. The design of the circular SiPM array is flexible, allowing for the size to be changed, which allows for the replacement of PMTs of different sizes. In particular, our design is advantageous in fields where there is a restriction on the size of the detectors, such as applications in space and handheld instruments. The issue of substantial thermal noise is the main hindrance to the application of SiPMs, and specifically, the signal-to-noise ratio is the key factor. In principle, the use of SiPMs is feasible for scintillation crystals with a high light yield and fast luminescence. Since the noise of the SiPMs mainly concentrates on individual electrons, it can be easily overcome by a relatively small threshold, and maintaining a low temperature is also an effective way to suppress noise. Another disadvantage of the large-size SiPM array is waveform broadening, which causes the waveform of the scintillator to be distorted. This is unfavorable for pulse-shape discrimination and time measurements. One solution is to develop high-speed front-end digitalization, and the best solution is to develop digital SiPMs in the near future.

**Acknowledge**

This research was supported by the Key Research Program of Frontier Sciences, Chinese Academy of Sciences (QYZDB-SSW-SLH012), the National Natural Science Foundation of China (11775251, 11775252), the strategic leading science and technology program of Chinese Academy of Sciences (XDA 15360100, XDA 15360102).